\documentclass[10pt]{article}
\usepackage{amsmath,amsfonts,amssymb,amsthm,bbm}
\usepackage{pdfsync}
\usepackage{graphicx}
\usepackage[latin1]{inputenc}
\usepackage{hyperref}
\usepackage[all]{xy}
\usepackage{stmaryrd}
\usepackage{authblk}

\newcommand{\bit}{\begin{itemize}}
\newcommand{\eit}{\end{itemize}}
\newcommand{\bd}{\begin{description}}
\newcommand{\ed}{\end{description}}

\newcommand{\bc}{\begin{center}}
\newcommand{\ec}{\end{center}}
\newcommand{\Ref}[1]{(\ref{#1})}

\newcommand{\C}{{\mathbb C}}

\newcommand{\R}{{\mathbb R}}
\newcommand{\Z}{{\mathbb Z}}
\def\T{\mathbbm T}

\newcommand{\cE}{{\mathcal E}}
\newcommand{\cF}{{\mathcal F}}
\newcommand{\cG}{{\mathcal G}}

\newcommand{\SU}{\mathrm{SU}}

\newcommand{\SL}{\mathrm{SL}}

\newcommand{\hh}{{\cal H}}

\newcommand{\be}{\begin{equation}}
\newcommand{\ee}{\end{equation}}
\newcommand{\bea}{\begin{eqnarray}}
\newcommand{\eea}{\end{eqnarray}}
\newcommand{\bs}{\begin{subequations}}
\newcommand{\es}{\end{subequations}}
\newcommand{\nn}{\nonumber}

\newcommand{\w}{\wedge}

\newcommand{\tr}{{\rm Tr}}
\newcommand{\f}{\frac}
\newcommand{\tl}{\tilde}
\def\p{\partial}
\newcommand{\Id}{\mathbbm{1}}

\newcommand{\re}{\mathrm{Re}}
\newcommand{\im}{\mathrm{Im}}

\newcommand{\ra}{\rangle}
\newcommand{\la}{\langle}
\newcommand{\bra}[1]{\langle {#1}|}
\newcommand{\ket}[1]{|{#1}\rangle}

\renewcommand{\a}{\alpha}  \newcommand{\g}{\gamma}
\renewcommand{\d}{\delta}  \newcommand{\eps}{\epsilon}  \newcommand{\z}{\zeta}
 \renewcommand{\th}{\theta}      \renewcommand{\l}{\lambda}
\let\m=\mu  \let\n=\nu   \newcommand{\s}{\sigma}     \let\vphi=\varphi  \let\om=\omega
\let\G=\Gamma   \let\Th=\Theta  \let\Si=\Sigma

\newcommand{\Ket}[1]{ \left| #1 \right\rangle}
\newcommand{\Bra}[1]{ \left\langle #1 \right|}
\newcommand{\Scal}[2]{ \left\langle #1 | #2 \right\rangle}
\newcommand{\KetQ}[1]{ \left| #1 \right]}
\newcommand{\BraQ}[1]{ \left[ #1 \right|}
\newcommand{\ScalQ}[2]{ \left[ #1 | #2 \right]}
\newcommand{\Miss}[2]{ \left[ #1 | #2 \right\rangle}
\newcommand{\MissQ}[2]{ \left\langle #1 | #2 \right]}

\newcommand{\norm}[1]{\bra{#1}#1\ra}
\newcommand{\po}{\pi\om}%{[\pi\ket{\om}}
\newcommand{\tpo}{\tl\pi\tl\om}%{[\tl\pi\ket{\tl\om}}

\newcommand{\Pois}[2]{ \left\{ #1 , #2\right\} }

\begin{document}

\title{\bf A note on the secondary simplicity constraints \\ in loop quantum gravity}

%\author{\Large{Fabio Anz\`a and Simone Speziale}
%\smallskip \\
%\small{Centre de Physique Th\'{e}orique, CNRS-UMR 7332, Aix-Marseille \& Toulon Universities, }
%\linebreak \\ \small{Luminy Case 907, 13288 Marseille, France}}
%\date{\today}

\author[1]{Fabio Anz\`a\thanks{fabio.anza@physics.ox.ac.uk}}
\author[2]{Simone Speziale\thanks{simone.speziale@cpt.univ-mrs.fr}}
\affil[1]{Atomic and Laser Physics, Clarendon Laboratory, University of Oxford\\

 Parks Road, Oxford, OX1 3PU, UK }
\vspace{1cm}
\affil[2]{Centre de Physique Th\'{e}orique, CNRS-UMR 7332, Aix-Marseille \& Toulon Universities\\

Luminy Case 907, 13288 Marseille, France}

\maketitle

%----------------------------------------------------------------------------
\begin{abstract}
\noindent 
A debate has appeared in the literature on loop quantum gravity and spin foams, over whether the secondary simplicity constraints, reducing the connection to be Levi-Civita, should imply the shape matching conditions, reducing twisted geometries to Regge geometries. We address the question using a simple model with a flat dynamics, in which secondary simplicity constraints arise from a dynamical preservation of the primary ones. We find that shape matching conditions arise, thus providing support to an affirmative question. The origin of these extra conditions is to be found in the different graph localisation of the Hamiltonian and primary simplicity constraints. Our results are consistent with previous claims by Dittrich and Ryan, and extend their validity to Lorentzian signature and arbitrary cellular decompositions.  We show in particular how the (gauge-invariant version of the) twist angle $\xi$ featuring in twisted geometries equals on-shell the Regge dihedral angle multiplied by the Immirzi parameter, thus recovering the discrete extrinsic geometry from the Ashtekar-Barbero holonomy. Finally, we confirm that flatness implies both the Levi-Civita and the shape-matching conditions using twisted geometries and a 4-dimensional version of the vertex condition appearing in 't Hooft's polygon model.
\end{abstract}
%----------------------------------------------------------------------------

%----------------------------------------------------------------------------
\section{Introduction}
%----------------------------------------------------------------------------

In this paper we are concerned by the possible presence and meaning of secondary simplicity constraints in the dynamics of loop quantum gravity, in particular in its definition by the spin foam formalism.
The current state of the art in spin foams is the EPRL model \cite{LS, EPR,FK,LS2,EPRL,KKL,CarloGenSF}, in which secondary constraints are not explicitly posited.
The question is rather central \cite{AlexandrovCritical}, and has received a certain amount of attention in the literature \cite{DittrichRyan,DittrichRyan2, AlexandrovNewVertex, Gielen:2010cu,Geiller:2011aa,BaratinOriti,HaggardSpin,Alexandrov:2012pj,WielandHamSF}.
In the continuum theory, these constraints arise from the stabilisation of the primary simplicity constraints. While the latter impose that the bivector $B^{IJ}$ is simple and thus expressed in terms of the tetrads, the former encode (the spatial projection of) Cartan's equation. They thus imply that the spacetime connection is Levi-Civita, and capture the non-linearity of Einstein's theory in the first order formalism.
The reason why the secondary constraints are usually not included is partly because their non-linearity introduces ambiguities and difficulties when discretising them, but mostly because of a key result due to Barrett and Crane \cite{BarrettCrane}. They discretised the covariant constraints on a flat 4-simplex, and split them into primary or secondary according to whether they involve only bivectors on adjacent boundary faces, or bivectors on opposite faces and the bulk connection. Then using classical geometry, they proved that gauge invariance (in the guise of the local closure conditions) plus the primary simplicity constraints are enough to impose also the secondary ones. This result has led to what could be called the `Barrett-Crane logic': imposing the primary constraints `at all times' (that is, on the full boundary of the 4-simplex) imposes automatically the secondary ones as well. This is a solid result, and indeed the EPRL model benefits from it, for instance it is crucial for the derivation of the Regge action at the saddle point approximation at large areas \cite{BarrettLorAsymp}, one of the many good properties of the model.
There are nonetheless unresolved issues, which justify the concerns of \cite{AlexandrovCritical} and the active research pursued in this area. First of all, the result by Barrett and Crane is classical, so quantum fluctuations in the path integral may not be properly taken into account. And even at the classical level, the result uses heavily the rigidity of a flat 4-simplex, and does not extend to arbitrary polytopes dual to general spin foams. General spin foams are however rather natural to define \cite{KKL,CarloGenSF}, and indeed necessary if one wants amplitudes for arbitrary loop quantum gravity states. Finally, there is evidence that on a large triangulation, the equations of motion defining the saddle point are only compatible with flat solutions, where the Regge deficit angle vanishes at each hinge \cite{HellmannFlatness}. These questions motivate the improvement of our geometric understanding of loop quantum gravity on a fixed graph. 

From the canonical viewpoint, the question has been addressed first by Dittrich and Ryan in \cite{DittrichRyan,DittrichRyan2,Dittrich:2012rj}. These authors used a Regge-like parametrization of the BF phase space, and explicitly discretised the torsion-less condition by requiring that the discrete connection is Levi-Civita, in the sense of being compatible with the discrete metric defined by the data (see also \cite{Zapata:1996ij}). Among a number of interesting results, their analysis highlighted the result that not all bivector configurations allow for a Levi-Civita connection, but only those satisfying the shape-matching conditions that reduce the twisted geometries parametrising the kinematical phase space to Regge geometries \cite{DittrichSpeziale,twigeo,IoPoly,IoCarloGraph}. This is somewhat surprising: in the continuum, the secondary constraints provide a restriction to the connection degrees of freedom only, and should not have anything to do with the matching of shapes, which is a property of the intrinsic geometry of space. Indeed, the claim that the secondary constraints should additionally imply the matching of shapes has been questioned in \cite{HaggardSpin}, where the authors prove that a suitable discretisation of the spatial torsion-less condition can be solved without imposing any shape matching, thus introducing a useful notion of Levi-Civita connection for twisted geometries. 
However, the discretisation used in the holonomy-flux variables mixes connection and metric degrees of freedom, so the interpretation of the secondary constraints in terms of purely connection variables is not guaranteed. Furthermore, any discretisation brings in a certain amount of arbitrariness, therefore it may be understandable that different conclusions have been reached in \cite{DittrichRyan} and \cite{HaggardSpin}. Since the discussion has consequences for the spin foam formalism, where usually one sums over histories of covariant twisted geometries, and not Regge geometries, it is important to clarify this debate. 
In order to shed light on the matter, we propose to consider a tighter model, where the secondary constraints are not posited by hand discretising one or another form of their continuum counterpart, but properly derived as stability equations for the primary simplicity constraints, considered here in the linear version introduced in \cite{EPRL}. 

In setting up such a dynamical derivation, one has to face two immediate difficulties. Firstly, in spin foam models one does not have a standard Hamiltonian picture with a space-like boundary phase space and continuous time. The phase space is associated with a general boundary enclosing a compact region of spacetime, and spacetime itself is discretised by gluing together flat patches, typically 4-simplices. 
Secondly, a fixed graph structure a priori breaks diffeomorphism invariance, so it is not obvious in which sense a precise classical Hamiltonian constraint can be defined on a fixed graph. 
These are core issues extensively discussed in the literature, see for instance \cite{Immirzi94, Zapata:1996ij, ThiemannAQG1, Bahr:2009ku, DittrichDiffeoLattice, DittrichHoehn, DittrichBonzom, Hoehn:2014fka}. However here we wish to focus on the more specific issue of whether secondary simplicity constraints can in principle be dynamically generated in a discrete setting, and if so, what is their meaning. 
To that end, we set up a model where this question can be studied bypassing the broader ones, by choosing the case in which diffeomorphism invariance can be implemented on a fixed lattice, that is the case of a flat spacetime. Namely, we take a spatial triangulation with continuous time, and choose a simple Hamiltonian leading to flatness of the 4-dimensional connection.

Our first result is to show that even such simple Hamiltonian leads to dynamically generated secondary constraints. These come from the stabilisation of the diagonal, Lorentz-invariant primary simplicity constraints. The two form a second class pair, precisely as in the continuum. Since the orbits of the diagonal simplicity constraints had been shown in \cite{IoWolfgang} to arbitrarily change the extrinsic curvature of the discrete twisted geometry associated to the graph, the breaking of this gauge symmetry is a positive fact. 
Crucially, there are more secondary constraints than primary: this is a consequence of the graph structure, in particular the fact that while the primary simplicity constraints are defined on the links, the Hamiltonian is defined on the faces. This implies that the secondary are not simply relations between the dihedral angle on each link and the SU(2) data, but also restrict the SU(2) data. Using the geometric description provided by twisted geometries, we are able to show that $(i)$ the additional restrictions correspond precisely to the shape matching conditions, and $(ii)$ these being satisfied, the covariant holonomy solving the secondary constraints is Levi-Civita in the sense of Regge, by relating 4-dimensional and 3-dimensional dihedral angles via the spherical cosine law. This shows in particular how to extract the extrinsic geometry from the reduced SU(2) Ashtekar-Barbero holonomy.
These results are in agreement with those of Dittrich and Ryan, notably with the claim that solving secondary constraints requires shape matchings, and extends them to Lorentzian signature and partially to arbitrary graphs, where the main limitation comes from the non-trivial adjacency matrix of polyhedra. However, 
they are reached within flat dynamics, and the simplicity of our model does not allow to state whether they would hold with more general Hamiltonians allowing for 4-dimensional curvature. Indeed, we also show that the shape matching conditions arise equally well from considering the Hamiltonian alone, and asking for non-degeneracy of the solutions, as they do from the secondary constraints alone. Therefore, all we can truly conclude is that a fully constrained system leading to flat dynamics has both secondary constraints and shape matching conditions. To further elucidate the relation between flatness, Levi-Civita holonomy and shape matching conditions, we use twisted geometries to generalise the vertex condition of 't Hooft's polygon model \cite{'tHooft} to four dimensions, and confirm that flatness implies both the Levi-Civita and the shape-matching conditions.

Throughout the paper we will use the spinorial and twistorial formalisms \cite{twigeo,twigeo2,EteraSU2UN,EteraSpinor,EteraTamboSpinor,WielandTwistors,IoHolo,IoTwistorNet,IoWolfgang,Livine:2013zha,IoNull,Livine:2013tsa,IoCreta}.
It allows us to perform canonical calculations using Darboux coordinates in phase space, instead of the non-commuting fluxes, and to bridge easily between structures which are local on the links, and structures which are local on the nodes. Specifically, we will take advantage of the relation between the Ashtekar-Barbero and Lorentzian holonomies established in \cite{IoWolfgang}, and of the geometric description of SU(2) invariants at the node proposed by Freidel and Hnybida in \cite{FreidelJeff13}, allowing to describe fully reduced variables for the gauge-invariant phase space, in particular, a gauge-invariant analogue of the twist angle $\xi$ of twisted geometries. 
We use $I=0\dots 3$ for Minkowskian  indices, $A=0,1$ for spinorial indices. 
We take mostly-plus signature, so spacetime vectors are mapped to anti-hermitian matrices. 
We label links of the graph by $i,j,k$, and try to avoid writing explicitly internal SU(2) indices. We use `triangulation' to refer to a simplicial discretisation of a 3-dimensional manifold, and `cellular decomposition' for more general discretisations where arbitrary polyhedra are allowed, ad not just tetrahedra. We refer to the individual structures of a cellular decomposition as vertices, edges, polygonal faces and polyhedra. A graph dual to a cellular decomposition has $N$ nodes corresponding to the polyhedra, $L$ links corresponding to the polygons, and $F$ faces  corresponding to the edges.

%----------------------------------------------------------------------------
\section{Simplicity constraints and gauge orbits}
%----------------------------------------------------------------------------

In the continuum theory, the secondary simplicity constraints turn the primary ones from first class to second class constraints. Accordingly, the solution to the secondary constraints can be interpreted as a non-trivial gauge fixing of the orbits generated by the primary constraints. Such gauge fixing plays a crucial role: it says that the spatial part of the connection is Levi-Civita, and this in turns allows to extract the extrinsic geometry from the Ashtekar-Barbero connection. This is a key property of the structure of the theory, and understanding how it is to be implemented on a fixed graph is necessary for the study of spin foam amplitudes. Fixing a graph corresponds to a truncation of the theory \cite{IoCarloGraph}, and introduces some peculiarities with respect to the continuum theory. Firstly, some of the primary simplicity constraints are already second class by themselves: this is a well-known consequence of the non-commutativity of the discrete fluxes. 
Secondly, the connection per se can not be properly spoken of: all it exist on a fixed graph is the holonomy. Therefore the embedding we are talking about is that of the SU(2) holonomy (appearing e.g. as the argument of spin network states) in the covariant phase space $T^*\SL(2,\C)^L$, associated to a graph with $L$ links.
To understand the details of these structures, let us briefly recall the twistorial representation of the Lorentzian holonomy-flux algebra, referring the reader to the literature for further details.
For this, we focus first on a single link of the graph.
The holonomy and self-dual generators for source ('untilded') and target ('tilded') nodes of a link are  
\be\label{hfdef}
\Pi^{AB}  = -\f12\om^{(A}\pi^{B)}, \qquad
h^A{}_B = \f{\tl\om^A \pi_B - \tl\pi^A\om_B}{\sqrt{\po}\sqrt{\tpo}}, \qquad 
\tl\Pi^{AB}  = \f12\tl\om^{(A}\tl\pi^{B)},
\ee
and the (complex) area matching constraint $C=\po-\tpo=0$ has to be satisfied in order for the twistor space $\T^2$ to describe $T^*\SL(2,\C)$ and its Poisson algebra. Here $\po:=\pi_A\om^A$ is the Lorentz invariant contraction, and we follow the conventions of \cite{IoWolfgang} and set $\{\pi_A, \om^B\} = \d_A^B = - \{\tl\pi_A, \tl\om^B\}$.
We work in the time gauge, where the time-like normal to each node is $n^I=(1,0,0,0)$. This allows us to identify the function on phase space corresponding to the external dihedral angle between the frames associated to the source and target nodes, that is
\be\label{defXi}
\cosh\Xi := -\tl{n}_I \Lambda(h)^{I}{}_{J} n^J, \qquad \Xi=2\ln \left(\frac{\|\om\|}{\|\tl{\om}\|}\right).
\ee
The linear simplicity constraints \cite{EPRL} in the time gauge read
\be \label{simpl}
\Pi^{AB} = e^{i\th} \Pi^\dagger{}^{AB}, 
\ee
where $\g = \cot \f\th2$ is the Immirzi parameter. They are to be imposed on both source and the target generators, and as the norms of source and target generators coincide, this amounts to 5 real constraints per link.
They can be conveniently split into a real, Lorentz-invariant diagonal constraint $S$, and two complex, $n^I$-dependent off-diagonal constraints $F$ and $\tl F$, 
\be
S = (\re-\g\im) \f{(\po)+(\tpo)}2=0, \qquad F = \d_{A\dot A} \pi^A \bar\om^{\dot A}=0,
\qquad \tl F = \d_{A\dot A} \tl\pi^A \bar{\tl\om}^{\dot A}=0.
\ee
The diagonal constraint is solved by requiring that $\po=\tpo=(\g+i)j$, with $j\in\R$, whereas the $F$ constraints can be solved eliminating part of the $\pi^A$ spinors in favour of the $\om^A$, and similarly for the tilded spinors.
 Only the diagonal constraint is first class, whereas the real and imaginary parts of $F$ form a second class pair,
\be\label{FFbar}
\{F, \bar F\} = 2 i \im(\po) \approx 2ij,
\ee
and equally for $\tl F$.
As anticipated above, the fact that some of the primary constraints are already second class by themselves is a peculiarity of the discretisation and the associated non-commutativity of the fluxes. Since we are interested in the non-trivial gauge fixing that would be provided by the secondary constraints, we focus attention on the orbits generated by the first class, 
diagonal constraint. They preserve the fluxes but change the holonomy,
\be\label{Sonh}
\{S, \Pi^{AB}\} =0, \qquad \{ S, h^A{}_B \} = -\f{1+i\g}2 \hat h^A{}_B, 
\ee
where
\be
\hat h^A{}_B  := \f{\tl\om^A \pi_B + \tl\pi^A\om_B}{\sqrt{\po}\sqrt{\tpo}} = 
-\f 4{\po} (h\Pi)^A{}_B.
\ee
As pointed out in \cite{IoWolfgang}, the key property of these orbits is to shift the dihedral angle $\Xi$: we have
\be
\{S, \Xi\}= 1,
\ee
hence $\Xi$ is pure gauge with respect to the primary constraints. Given its interpretation in terms of the 4-dimensional dihedral angle, this result is already a strong indication that secondary constraints are mandatory if we want to be able to reconstruct the extrinsic geometry out of the holonomy-flux phase space: if, as in the continuum, the secondary constraints turn the primary into second class, their solution will provide a non-trivial gauge-fixing of the orbits, that is a relation between $\Xi$ and the reduced data.

Such reduced data can be found in two steps \cite{IoWolfgang}; first, reducing $\T^2$ by the simplicity constraints (4 second class and 1 first class), giving two spinors satisfying the reduced area matching condition,
\be\label{zred}
z^A = \sqrt{2j}\frac{\om^A}{\|\om\|^{i\g+1}}, 
\qquad  \tl z^A = \sqrt{2j}\frac{\tl\om^A}{\|\tl\om\|^{i\g+1}}, \qquad C=\|z\|^2-\|\tl{z}\|^2=0,
\ee
where $\|z\|^2=2j$, and induced Poisson brackets $\{z^A, \bar{z}^{\bar B}\} = i \d^{A\dot B} = - \{\tl z^A, \bar{\tl z}^{\bar B}\}$.
Then, reducing by the first class constraint $C$, a step which gives the symplectic manifold $T^*\SU(2)$, as shown in \cite{twigeo2},
\be\label{T*SU2}
\vec X = \f12\bra{z}\vec\s\ket{z}, \qquad 
 g = \f{\ket{\tl z}\bra{z}+|\tl z][z|}{\|z\| \|\tl z\|},
\qquad \vec {\tl X} = -\f12\bra{\tl z}\vec\s\ket{\tl z},
\ee 

What is new thanks to the Lorentzian starting point is the possibility to appreciate the role of the Immirzi parameter $\g$: it introduces a phase shift in the relation between left-handed and right-handed structures \Ref{simpl}, captured by the $\g$-dependent phase of the reduced spinors \Ref{zred}. This in turn twists the reduced holonomy $g$ in such a way that it is not anymore the restriction of $h$ to the stabiliser of $n^I$. Precisely as in the continuum, the twist introduces a dependence of the reduced holonomy on the boosts degrees of freedom of the Lorentz holonomy.
This can be made explicit if we reintroduce the orbits of $S$, and look at the reduced space as if $S$ were second class. This space has the structure of a bundle $T_\Xi :=T^*\SU(2)\times \R$, where the fibres are the orbits of $S$. On $T_\Xi$, the Lorentzian holonomies and fluxes reduce to
\be\label{hf-red}
\Pi^{AB} \approx -\f12 (\g+i) z^{(A} \d^{B)\dot B}\bar z_{\dot B}, \qquad 
h^A{}_B \approx \f{e^{-(i\g+1)\f\Xi2}\ket{\tl z}\bra{z}+e^{(i\g+1)\f\Xi2}|\tl z][z|}{\|z\| \|\tl z\|},
\ee
which makes it clear that the relation between the matrices $g$ and $h$ mixes boosts and rotations for non-vanishing $\g$. In the following, we will often suppress the matricial indices for ease of notation.
Using an interaction picture to factorise the path-ordered exponentials defining the holonomies, one can prove \cite{IoWolfgang} that 
\be\label{gh}
g = h V_K^{-1} V_K^\g,
\ee
where 
\be
V_K = \f{e^{-\Xi/2} \ket{z}\bra{z}+ e^{\Xi/2}|z][z|}{\norm{z}^2}, \qquad 
V_K^\g = \f{e^{i\g\Xi/2} \ket{z}\bra{z}+ e^{-i\g\Xi/2}|z][z|}{\norm{z}^2}. \qquad 
\ee
The relation \Ref{gh} is the fixed-graph version of the well-known relation between the antiself-dual Lorentz connection $\om_{ASD}$, and the Ashtekar-Barbero connection, $A_{AB} = \om_{ASD} + (\g-i)K$, where $K$ is the boost part, and shows that $g$ should be thought of as the lattice version of the Ashtekar-Barbero connection.
This is a neat result, that in our view shows the usefulness of the twistorial formalism for loop gravity.
The purpose of this paper is to study how the dynamics introduces a non-trivial gauge-fixing of the orbits, which in turn allows to extract directly the extrinsic geometry out of the reduced data.

%----------------------------------------------------------------------------
\section{Secondary constraints from flat dynamics}
%----------------------------------------------------------------------------

We take a hybrid discretisation approach, whereby we foliate spacetime as $M=\Sigma\times\R$, where $\Si$ is discretised via an arbitrary cellular decomposition dual to the graph $\G$, and $t\in\R$ is a continuous time parameter. This is the same approach used for instance in \cite{Zapata:1996ij,DittrichRyan} and it allows for a straightforward construction of a symplectic form on phase space.\footnote{Alternatively, one can discretise the full spacetime manifold, like in the spin foam formalism. In this case constraints can be defined in a covariant language looking for instance at symmetries of the action, see e.g. \cite{DittrichHoehn} for this approach. A hybrid set-up has been proposed in \cite{WielandHamSF}, with discretised spacetime, and a continuous `time' parameter associated to the bulk faces of the 2-complex. Our results could easily extend to these other approaches, as we will comment upon in the conclusions.}
The kinematical phase space $S_\G$ associated to the graph $\G$ is defined by a copy of $T^*\SL(2,\C)$ on each link, described by simple twistors as above, plus the Gauss law imposing gauge invariance at each node. Because we are working in the time gauge, boosts are fixed and the Gauss law is restricted to the SU(2) stabiliser of $n^I$, that is $\vec G_n=\sum_{l\in n}\vec L_l$.
The inclusion of this SU(2) closure constraint does not change the classification of the area-matching and simplicity constraints, as it commutes with both. The restriction of the graph to be dual to a triangulation means that we also have a notion of faces of the graph, bounded by links, which are dual to the edges of the cellular decomposition.

The system is very similar to the one considered in \cite{DittrichRyan}, where additional constraint were added, coming from a suitably discretised form of the torsion-less condition of the continuum theory. These constraints are then shown to contain the shape-matching conditions reducing twisted geometries to Regge geometries. As discussed in the introduction, this specific claim has been disputed by other authors \cite{HaggardSpin}, and it is our propose here to shed light on the debate by deriving the secondary constraints from a suitable discrete dynamics. To that end, we consider the following Hamiltonian constraint, 
\be\label{defH}
\hh_f = \im \tr \, h_f =0, \qquad h_f := \vec\prod_{l\in f} h_l,
\ee
for all elementary faces $f$ of the graph, which implies some partial flatness of the graph.

Adding Lagrange multipliers for all the constraints, the full action reads
\be\label{S}
S = \int dt \sum_{l} \pi_{l}\dot\om_l-\tl\pi_{l}\dot{\tl\om}_l 
+ \left[\n_l C_l  +\m_l F_l+\tl\m_l \tl F_l + cc.\right] + \l_l S_l + \sum_f N_f \hh_f + \sum_n \vec n_n \cdot \vec G_n.
\ee
Let us see how the addition of the Hamiltonian constraint changes the structure of the system.
${\cal H}_f$ commutes with itself, and with all the area matching and closure constraints. 
On the other hand, it does not commute with the simplicity constraints, because of the non-vanishing bracket between $S$ and the holonomy, see \Ref{Sonh}.
\begin{align}
\label{HD} & \{ \hh_f, S_{l}\} = -2(\g\re +\im) \left[\f1{\po_l}\tr\, (h_{f_l}\Pi_l)\right],
\end{align}
where we introduced the notation $h_{f_l}$ for the face holonomy starting at the source of the link $l$.

The complete canonical analysis is reported in Appendix A. For our purposes, we only need here the two main results that emerge from it. Firstly, assuming a non-vanishing lapse, the stability of $S_l$  leads to secondary constraints, which can be read immediately from the bracket \Ref{HD},
\be\label{sec}
\psi_{lf} := -2(\g\re +\im) \left[\f1{\po_l}\tr\, (h_{f_l}\Pi_l)\right]=0, \qquad \forall \, l\in\p f.
\ee
Secondly, the full system is over constrained, leading to a zero dimensional phase space, with a flat solution characterised by
\be
h_f =\Id,\quad \forall f,
\ee
and a first class Hamiltonian constraint generating shifts in the areas. 
This is to be expected, and fully consistent with previous analysis \cite{DittrichRyan}.

The key property of the secondary constraints is to turn the diagonal primary simplicity constraints into second class:
\begin{subequations}\label{psiS}\begin{align} \label{psiS1}
& \{S_l, \psi_{lf} \} \approx \f\g2\re\tr\, h_f \neq 0, \\
& \{S_l, \psi_{l'f} \} \approx 4\left(2\g \re +(1-\g^2)\im\right) \left[\f1{\po_l\po_{l'}} \tr\, \left(h_{f_l} \Pi_{l'}^{(l)}\Pi_l \right)\right], \qquad l'\neq l,
\end{align}\end{subequations}
where $\approx$ means on shell of the primary constraints, and $\Pi_{l'}^{(l)} := - h_l^{-1} \Pi_l h_l$ is the parallel-transport of $\Pi_{l'}$ at the vertex shared with $l$.
The equations \Ref{sec} and \Ref{psiS} are our first result: a simple Hamiltonian constraint, such as the partially flat condition \Ref{defH}, is already enough to generate secondary constraints, and turn the first class primary simplicity into second class. This is an encouraging result, because it implies that $\Xi_l$ do not drop out from the reduced phase space, and the latter therefore preserves the information about the extrinsic geometry. 
The next question is how to solve the secondary constraints and recover explicitly the extrinsic curvature out of the SU(2) phase space. Before addressing it, let us make a few remarks on the structure of the secondary constraints \Ref{sec} we derived:
\begin{itemize}
\item They can be taken as equations for the variables $\Xi_l$, providing a non-trivial gauge fixing for the orbits of the diagonal primary simplicity constraints, or in other words, a non-trivial section on the bundle $T_\Xi$ on each link.
\item They are non-linear, and not local in the sense of the graph, but almost: they involve first-neighbours of each pair link-face on which they are defined; hence, the solutions will relate such neighbouring data.
\item Crucially, there are more constraints than links, contrarily to the intuition from the continuum theory. This is due to the fact that the simplicity constraints are local on links, whereas the Hamiltonian is local on faces. As a consequence, each link has as many secondary constraints as there are faces sharing it, a fact that will have strong implications.
\item Like the primary simplicity, the secondary are conditions fixing the phase of a complex number. As in the continuum, they introduce a relation between the connection and the metric degrees of freedom; however, in the form \Ref{sec} they are not manifestly a discretisation of the torsion-less condition, but rather conditions on the Lie derivative of the loop holonomy, or, conversely, on the curvature seen by the flux vector. Nonetheless, we will see below that solving these constraints amounts exactly to the statement that the covariant connection is Levi-Civita, and so the torsion vanishes.
\item The expression \Ref{sec} also shows that, had we started with an Hamiltonian imposing complete flatness,\footnote{As was done in a previous version of this paper, by further imposing $\re\tr\,h_f=2$.} the secondary constraints would be redundant, and the system too simple for our purposes. With the weaker Hamiltonian \Ref{defH}, the secondary constraints are fully independent, and flatness only arises from the stability procedure as described above.

\end{itemize}

In order to proceed, it is convenient to manipulate the Hamiltonian and the secondary constraints and rewrite them using a basis of nodal $\SL(2,\C)$ invariants. Using spinors, these can be written as the following bilinears considered for instance in \cite{IoHolo},
\be\label{EFG}
{\cal E}_{ij} := \om_i\cdot\pi_j, \qquad {\cal F}_{ij} := \om_i\cdot\om_j,
 \qquad {\cal G}_{ij} := \pi_i\cdot\pi_j.
 \ee
The usual invariants given by scalar and mixed products of generators can be easily expressed in this basis, which furthermore form a closed algebra of Poisson brackets.\footnote{When all $N$ links have the same orientation with respect to the node, ${\cal E}_{ij}$ generate a $\mathfrak{gl}(N,\C)$ sub-algebra, and ${\cal F}_{ij}$ and $\bar{\cal G}_{ij}$ commute among each other. When the orientation of the link is mixed, one can use the map $(\om^A\mapsto \pi^A, \ \pi^A\mapsto -\om^A)$.}
For ease of notation we have omitted a node label in the above invariants, and as we have restricted attention to graphs with nodes connected by one link at most no ambiguity arises.
Consider a 3-valent face; expanding the trace of the loop holonomy using \Ref{hfdef} and \Ref{EFG}, we obtain
\begin{align}\label{3v-H}
{\prod_l \po_l}  \, \tr \prod_{l\in f} h_l = -\cE_{23} \cE_{12} \cE_{31} - \cE_{32} \cE_{21} \cE_{13} 
+ \cE_{23} \cF_{31} \cG_{12} +  \cE_{32} \cF_{12} \cG_{31} \\
+ \cE_{31} \cF_{12} \cG_{23} + \cE_{21} \cF_{31} \cG_{23} 
+  \cE_{12} \cF_{23} \cG_{31} +  \cE_{13} \cF_{23} \cG_{12}.
\end{align}
In this way, the the Hamiltonian and the secondary constraints equal (up to normalisation) 
cubic polynomials of invariants.

On-shell of the primary simplicity constraints, using \Ref{zred}, we have
\be
{\cal E}_{ij} \approx -\f{\g+i}2 \sqrt{\f{j_i}{j_j}} \left(\f{\|\om_i\|}{\|{\om_j}\|}\right)^{i\g+1} \bar E_{ij}
, \qquad {\cal F}_{ij}\approx \f{ \left( {\|\om_i\|}{\|{\om_j}\|}\right)^{i\g+1} }{2\sqrt{{j_i}{j_j}}} F_{ij}, 
\qquad {\cal G}_{ij} \approx \f{(\g+i)^2 \sqrt{ j_i j_j}}{2\left( {\|\om_i\|}{\|{\om_j}\|}\right)^{i\g+1}} \bar F_{ij},
\ee
where 
\be\label{defEF}
E_{ij}:=\bra{z_i}z_j\ra, \qquad F_{ij}:=[z_i\ket{z_j}
\ee
are a basis of SU(2) invariants. Introduced in this context in \cite{Girelli:2005ii}, they also form a closed algebra, with ${\frak u}(N)$ sub algebra generated by the $E_{ij}$. Notice that the $F_{ij}$ are not linearly-independent: they satisfy the Pl\"ucker identities 
\be\label{Plu}
F_{ij}F_{kl}=F_{ik} F_{jl} -F_{il}F_{jk}, 
\ee
as a direct consequence of the spinorial identity.\footnote{The fact that Pl\"ucker relations appear in the description of SU(2) invariants may appear puzzling at first, but as was pointed out in \cite{EteraSU2UN}, the phase space on each node is indeed a projective space, given by the Grassmannian ${U(N)}/{SU(N-2)\times U(2)}$.} 
Recalling the expression \Ref{defXi} for the dihedral angle $\Xi$, the reduced Hamiltonian and secondary constraints can be written in a compact form as follows,
\begin{subequations}\label{traces}\begin{align}\label{3v-Hred}
\tr \, h_f & \approx  \prod_l (2j_l)^{-1}
\sum_{\m=0}^3 \s_{\m} (e^{-\f i2(\g-i)\Th_\m} A_\m + e^{\f i2(\g-i)\Th_\m} \bar A_\m), \qquad \s_\m=(1,-1,-1,-1)
\\ \label{psi-red}
-\f4{\po_l}\tr\, (h_{f_l}\Pi_l) &\approx \prod_l (2j_l)^{-1} \sum_{\m=0}^3 \s_{\m l} \Big(e^{-\f i2(\g-i)\Th_\m} A_\m - e^{\f i2(\g-i)\Th_\m} \bar A_\m\Big), \qquad 
\s_{\m l} = \left\{\begin{array}{ll} 1 & \m=0, \ \m=l \\ -1 & \m\neq l \end{array}\right.
\end{align}\end{subequations}
where
\begin{align}
& \Th_0=\Xi_1+\Xi_2+\Xi_3, && \Th_1= -\Xi_1+\Xi_2+\Xi_3, && \Th_2=\Xi_1-\Xi_2+\Xi_3,
&& \Th_3=\Xi_1+\Xi_2-\Xi_3, \\
& A_0=\bar E_{12} \bar E_{31} \bar E_{23}, && A_1 =  \bar E_{23} F_{13} \bar F_{21}, 
&& A_2 =  \bar E_{31} F_{21} \bar F_{32},  && A_3 =  \bar E_{12} F_{32} \bar F_{13}.
\end{align}
Analogue expressions can be obtained for faces of arbitrary valence. 
In spite of their apparent complexity, we will now show that these constraints can be explicitly solved. To that end, we need to trade the spinors for more geometric variables. 

%----------------------------------------------------------------------------
\section{Twisted geometries and spinor phases}
%----------------------------------------------------------------------------

To improve the geometric understanding of the system, and be able to solve explicitly the constraints, it is convenient to trade the spinors for twisted geometry variables, representing the intrinsic and extrinsic geometry of the polyhedra associated with a cellular decomposition dual to the graph. 
The reduction from spinors to twisted geometries requires to perform the explicit reduction by the two sets of first class constraints corresponding to area-matching and closure. Since the constraints are associated to different structures of the graph, respectively to links and nodes, solving one set first or the other leads to different intermediate steps, and thus different parametrisations of the final result, see the following scheme.

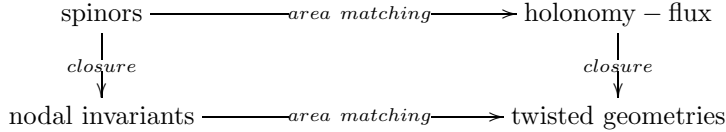
\begin{displaymath}
    \xymatrix@C=4cm{ {\rm spinors} \ar[r]|{  area \ matching}\ar[d]|{closure} & {\rm holonomy-flux} \ar[d]|{ closure}\\
       {\rm nodal \ invariants} \ar[r]|{area \ matching} & {\rm twisted \ geometries}}
\end{displaymath}

The `right-down' path involves solving first the area-matching reduction, and parametrizing the SU(2) holonomies and fluxes in terms of normal vectors and a twist angle,
\be\label{twigeo}
j=\f{\norm{z}}2, \qquad \xi = 2\arg z^1 - 2\arg\tl z^1, \qquad \z=\f{z^0}{z^1}, \qquad \tl\z=\f{\tl z^0}{\tl z^1}.
\ee
A nice feature of this parametrization is that $(j,\xi)$ forms an abelian conjugate pair. The Hopf sections $\z$ and $\tl\z$ provide stereographic coordinates for the sphere, with $N(\z)$ and $\tl N(\tl \z)$ the associated unit vectors, identified with the directions of the fluxes \Ref{T*SU2},
\be\label{defX}
\vec X = \f12 \bra{z} \vec\s \ket{z} = j N(\z), \qquad \vec {\tl X} = - \f12 \bra{\tl z} \vec\s \ket{\tl z} = - j \tl N(\tl\z)
\ee
Consider next the closure condition around each node imposed by gauge invariance. By Minkowski's theorem, these data define a unique convex polyhedron, whose adjacency matrix and intrinsic shapes are determined by the  scalar products $\vec X_i \cdot \vec X_j$, see \cite{IoPoly} for details. In particular, all properties of a polyhedron with $f$ faces are determined by the $f$ areas and $2f-6$ dihedral angles.
On the other hand, the twist angle $\xi_i$ between two adjacent frames is not gauge invariant, and in the reduction it has to be traded for a gauge-invariant angle built from the loop holonomies. This leads to cumbersome expressions which hinder an explicit solution of the constraints \Ref{psi-red}. The analysis simplifies considerably if we take the `down-right' path in the scheme, thanks to the results of \cite{EteraSU2UN, FreidelJeff13}, which we now review, based on the geometric interpretation of the nodal invariants as `framed' polyhedra. 

The idea is that the global  phase of the spinor, which is irrelevant to characterise $\vec X$, can be used to provide a framing vector $\vec F$ on the face \cite{EteraSU2UN}. This is defined as
 \be\label{defF}
 \vec F = \f1{2j} \im [z|\vec\s\ket{z},
 \ee
and satisfies $\vec F^2=1$, and $\vec F\cdot\vec X=0$.\footnote{The following inverse relations prove also useful,
\[
\ket{z}\bra{z} = j (\Id + \vec{N} \cdot \vec{\sigma}),
\qquad
\ket{z}[z| = j (\vec{F} \times \vec{N} + i \vec{F}) \cdot \vec{\sigma}.
\]
}
On a link, we have two such framing vectors, corresponding to source and target spinors. The angle between them can be immediately identified with the twist angle: 
in a common frame with both $N$ and $\tl N$ normals to the shared face along the $z$ direction, a straightforward calculation gives
\be\label{FFtl}
\vec F\cdot \vec{\tl F} = \cos\xi.
\ee
Then, let us follow \cite{FreidelJeff13} and parametrize the SU(2) spinor invariants as
\begin{subequations}\label{z-geo}\begin{align}
& \bra{z_i}z_j\ra = \sqrt{4 j_i j_j} \cos\f{\phi_{ij}}2 e^{-\f i2 (\a^i_j - \a^j_i)}, \\
& [z_i\ket{z_j} = \eps_{ij} \sqrt{4 j_i j_j} \sin\f{\phi_{ij}}2 e^{\f i2 (\a^i_j + \a^j_i)},
\end{align}\end{subequations}
where $\eps_{ij}$ is a sign introduced for commodity of parametrization.
These formulae will play a crucial role for solving explicitly the secondary constraints.

Using \Ref{defX} and \Ref{defF}, it can be easily shown that 
\be\label{defangoli}
\f{\vec X_i \cdot\vec X_j}{|\vec X_i| \, |\vec X_j|} = \cos\phi_{ij}, \qquad
\f{\vec X_i\w \vec X_j}{|\vec X_i\w \vec X_j|} \cdot \vec F_i = \cos \a^i_j, \qquad
\f{\vec X_j\w \vec X_i}{|\vec X_j\w \vec X_i|} \cdot \vec F_j = \cos \a^j_i,
\ee
which relates the angles appearing in \Ref{z-geo} to various scalar products. These formulas hold independently of the closure condition, as they just refer to the individual pairs of vectors. But when the closure holds we have an explicit geometric interpretation in terms of the polyhedron: for adjacent faces $(ij)$, $\phi_{ij}$ are the 3-dimensional external dihedral angles, $\vec X_i\w \vec X_j$ the edge vector, and $\a^i_j$ the angles between the edge vectors and the framing vectors. Taking scalar products between the edge vectors, one can immediately check that $\a^i_{jk}:=\a^i_j-\a^i_k$ are the 2d angles among the vectors associated with the edges $(ij)$ and $(ik)$ and thus, when they are adjacent, the 2d dihedral angles of the polygon. The adjacency matrix, key ingredient of this geometric interpretation, can be computed from the normals $\vec X_i$ using the reconstruction algorithm given in \cite{IoPoly}.\footnote{See also \cite{HaggardPentaVolume} for an analytic treatment of the 5-valent case.}
Finally, notice that there is a $2-to-1$ map between spinor configurations and reconstructed geometries, coming simply from the usual extra sign when describing a spinor in terms of its flagpole and flag. As spinors' phases, the $\a^i_j$'s are defined modulo $2\pi$, but the reconstructed geometry \Ref{defangoli} is modulo $\pi$.\footnote{On top of this $\Z_2$ symmetry, there is another subtlety that should be kept in mind when applying \Ref{z-geo} in practical calculations. It is customary to take both normals on a link pointing outwards in their respective frames. This introduces a parity transformation, so to have the minus sign in the adjoint action $\tl X = - g X g^{-1}$.
Twisted geometries can be parametrised with the parity either in the holonomy, as in the original paper \cite{twigeo}, or in the generators, as we did here in \Ref{T*SU2}. This is convenient to parametric in the same way for the source and target spinors the solution to the simplicity constraints. 
As a consequence, the `tilded' outgoing normal is given by the parity transformed spinor $|z]$, see \Ref{defX} -- recall that $[z|\s|z] = -\bra{z}\s\ket{z}$. In terms of the angles, this corresponds to 
\be\nn
{\cal P}: \qquad \ket{z_i} \mapsto |z_i] ,\qquad 
\Leftrightarrow \qquad \big(\phi_{ij} \mapsto \pi-\phi_{ij}, \ \a^i_j \mapsto -\a^i_j+(1-\eps_{ij})\pi, \ \a^j_i \mapsto \a^j_i\big).
\ee
}

To evaluate the brackets between the $\a^i_j$ and the other canonical variables, we plug \Ref{twigeo} in \Ref{z-geo}, getting the relation 
\begin{align}
& \a^i_j = 2\arg z_i^1 + \vphi^{+-}_{ij} - \vphi^{--}_{ij} + \f{1-\eps_{ij}}2\pi \\
& \a^j_i = 2\arg z_j^1 + \vphi^{+-}_{ij} + \vphi^{--}_{ij} + \f{1-\eps_{ij}}2\pi, 
\end{align}
where $\vphi^{AB}_{ij}$ is the phase of the matrix element $\bra{A}n^\dagger(\z_i) n(\z_j)\ket{B}$.
Then, it is easy to compute
\be\label{PBaj}
\{\a^i_j, j_i\} = 1, \qquad \{\a^i_j, j_j\} = 0.
\ee
For incoming links, which have spinorial Poisson brackets of opposite sign, we will use a notation $\a^{\tl \imath}_j$, and
\be\label{PBajtl}
\{\a^{\tl \imath}_j, j_i\} =-1, \qquad \{\a^{\tl \imath}_j, j_j\} = 0
\ee
The $\a^i_j$'s are thus conjugated to the areas, albeit do not form canonically conjugated pairs, since the phases $\a^i_j$ are not linearly independent: they satisfy the Pl\"ucker identities induced from \Ref{Plu}. 
Notice also that they are not invariant under the action of the area-matching constraint: from \Ref{PBaj} and \Ref{PBajtl} we immediately see that 
\be
\{C_i , \a^i_j\} = -2 = \{C_i, \a^{\tl \imath}_k\}.
\ee
Hence, any phase difference of the type
\be
\xi^i_{jk} := \a^i_j-\a^{\tl \imath}_k
\ee
is area-matching invariant. And as the $\a^i_j$ are rotational invariant at the nodes, these angles are invariant under both closure and the area matching constraints, and provide good variables for the fully reduced phase space, as already pointed out in \cite{FreidelJeff13}.

Replacing $\xi_i$ by $\xi^i_{jk}$ solves the issue of the non-gauge invariance of the former, and it is a neat improvement in the understanding of twisted geometries. 
The drawback of working with the $\xi^i_{jk}$ is that there is too many of them, one for each face sharing the link, and one should select a single representative for each link. 
Once this choice is made, the gauge-invariant reduced phase space of dimensions $2L+2N$ can be parametrised by the areas $j_l$, the chosen representative $\hat{\xi}{}^i_{jk}$, and the intrinsic shape variables for each polyhedron (2 for the tetrahedron, in general $2f-6$ for a polyhedron with $f$ faces), functions of the $\phi$'s.
Notice that we have arrived at a construction completely analogous to the one proposed for a triangulation by Dittrich and Ryan \cite{DittrichRyan}, where one defines multiple `proto-dihedral' angles on each link, and chooses one (say the averaged sum) as the phase space variable.

As the reader familiar with twisted goemetries knows well, this phase space describes a discontinuous metric, where the face shared by two polyhedra has a unique area but different shapes in the two frames. The shape-matching conditions making the metric Regge-like have been studied in \cite{DittrichSpeziale} for a triangulation, and \cite{IoPoly} for arbitrary polyhedra. For a triangulation, these can be characterised in terms of the $\a^i_j$ phases, as the matching of the 2-dimensional dihedral angles seen from the two adjacent frames. In our notations, 
\be\label{aSM}
\a^i_j - \a^i_k = \a^{\tl\imath}_l - \a^{\tl\imath}_m,
\ee
where $jil$ and $kim$ are links belonging to two different faces. Notice for later purposes that this implies that the gauge-invariant angle $\xi^i_{jk}$ is independent of the lower labels $(jk)$, that is it is unique for a given face. 
Imposing such condition leads precisely to a Regge geometry, which can be parametrised by dihedral angles on the triangles dual to the links, and lengths of the edges dual to the faces. This can be directly seen by construction, see also \cite{DittrichRyan}, and confirmed by an explicit counting: there are 2 shape-matching conditions per triangle, with one redundancy per edge, thus the reduced space has dimensions $2L+2N-(2L-F) = L+F$ matching thus of Regge calculus. Here we used the condition $L=2N$ valid for a closed triangulation.

For an arbitrary cellular decomposition, only some of the angles $\a^i_{jk}$ defined by the scalar products are true dihedral angles, thus imposing the matching \Ref{aSM} for all of them is redundant. Furthermore, matching the dihedral angles is not enough to make the metric continuous. A polygon with $p$ sides has $2p-3$ degrees of freedom, that can parametrised by the lengths of its $p$ sides and its $p-3$ diagonals, or in variables better suited to our purposes, by its area, its $p-1$ dihedral angles, and its $p-3$ diagonals. Thus, one has to match the dihedral angles but also the diagonals. The latter can be written has sums of consecutive edge vectors, thus it is not hard to write explicitly such generalised shape-matching conditions. Identifying the true dihedral angles and diagonals is of course complicated because in our phase space the adjacency matrix of a given polyhedron, and the thus the valency of each of its faces, varies as we explore the phase space. This variability, studied in details in \cite{IoPoly}, includes all possible changes of the boundary of a polyhedron by $2-2$ Pachner moves, but also reductions in the number of edges, as an edge can have zero length without leading to a degenerate metric, but simply to a face of smaller valence. Nonetheless, it is easy to provide an estimate of the global number of shape matching conditions on a graph, using average quantities. For a closed graph, each polyhedron will have on average $f=2L/N$ faces; and for polyhedra of dominant class, the average valence of its faces is $\la p\ra=6(1-2/f)$; counting $2p-4$ shape matching conditions per face, and removing like in the case of a triangulation a redundancy per edge, we obtain  
\be
\sum_l (2p_l-4) - F  \simeq \sum_l (2\la p\ra-4) - F = 8L-12N-F.
\ee
Hence, the dimensionality of the reduced space where the matching holds is
\be
6L-6N - (8L-12N-F) = 6N-2L+F.
\ee
It is interesting to remark that this number is \emph{smaller} than $L+F$ (recall that for a cellular decomposition the valence of each node is at least 4), which means that the reduced variables are not arbitrary edge lengths and face dihedral angles, as it would be if one naively would extend the Regge calculus logic from triangulations to polyhedral decompositions. We can instead take the independent variables to be $L$ dihedral angles $\th_l$ and $F-3L+6N\leq F$ edge lengths $\ell_e$. The fact that not all edge lengths are free is actually quite intuitive: the shape of a face of a polyhedron is determined by all the edge lengths in the polyhedron,\footnote{To be precise, the intrinsic geometry of 3-dimensional polyhedra is determined by areas and normals \cite{IoPoly}, however apart from configurations with a certain amount of regularity -- the typical example being a parallelepidedoid -- these can be inverted for the edge lengths.} therefore, when we fix one shape in order to match that of the adjacent polyhedron, we are imposing constraints on the edge lengths. 

In concluding this Section, let us summarise the various angle variables that have been introduced so far, and their geometric interpretation: 

\medskip

\begin{center}
\frame{\begin{tabular}{c|l} \\
$\xi_i = 2\arg z^1_i - 2\arg \tl z^1_i$ & twist angle \\ \\
$\a^i_j = \f i2\ln \f{\bra{i}j\ra}{\bra{j}i\ra} \f{\bra{j}i]}{[i\ket{j}}$ & closure-invariant spinorial phases \\ \\
$\xi^i_{jk}= \a^i_j-\a^{\tl \imath}_k$ & closure- and area-matching-invariant phases \\ \\
$\a^i_{jk} = \a^i_j-\a^i_k$ & 2d dihedral angle \\ \\
\end{tabular}}
\end{center}

%----------------------------------------------------------------------------
\section{Solving the secondary constraints}
%----------------------------------------------------------------------------

Now that we have indulged enough on the formalism, we are in a position to show how the secondary constraints can be explicitly solved, giving the discrete Levi-Civita connection.
We continue to consider the case of tri-valent faces, directly relevant for the 4-simplex, for the sake of being explicit. However the formulas immediately extend to arbitrary faces, and we will discuss the general case in the end.
As a first step, we substitute the geometric decomposition of the spinor bilinears \Ref{z-geo} in the $A_\m$ coefficients of \Ref{psi-red}. We obtain
\be
\prod_l (2j_l)^{-1} A_\m e^{-\f i2(\g-i)\Th_\m} = a_\m e^{-\f i2(\g\Th_\m -\chi_\m)-\f12\Th_\m}, 
\ee
where we introduced the shorthand notations 
\begin{subequations}\label{amu}\begin{align}
& a_0 =  \cos\f{\phi_{12}}2 \cos\f{\phi_{23}}2 \cos\f{\phi_{31}}2, &&
a_1 = \eps_{31}\eps_{12}\cos\f{\phi_{23}}2 \sin\f{\phi_{31}}2 \sin\f{\phi_{12}}2, \\
& a_2 = \eps_{12}\eps_{23}\cos\f{\phi_{31}}2 \sin\f{\phi_{12}}2 \sin\f{\phi_{23}}2,
&& a_3 = \eps_{23}\eps_{31}\cos\f{\phi_{12}}2 \sin\f{\phi_{23}}2 \sin\f{\phi_{31}}2,
\end{align}\end{subequations}
and
\be\label{chixi}
\chi_0:= \xi^1_{23} + \xi^2_{31} + \xi^3_{12}, \quad \chi_1 := - \xi^1_{23} + \xi^2_{31} + \xi^3_{12}, 
\quad \chi_2 := \xi^1_{23} - \xi^2_{31} + \xi^3_{12}, \quad \chi_3 := \xi^1_{23} + \xi^2_{31} - \xi^3_{12}.
\ee
As expected, only the geometric (that is, gauge-invariant and area-matching invariant) angles $\xi^i_{jk}$ enter the expressions.
The formulae \Ref{z-geo} allow us to write the traces, and thus the Hamiltonian and secondary constraints, as equations among trigonometric functions of dihedral angles:
\begin{subequations}\label{redconstraints}\begin{align}\label{H-red}
 \tr\, h_f & \approx 2  \sum_{\m=0}^3 \s_\m a_\m \left( 
\cos\f{\g\Th_\m -\chi_\m}2 \cosh \f{\Th_\m}2 + i \sin\f{\g\Th_\m -\chi_\m}2 \sinh \f{\Th_\m}2 \right), \\
 -\f4{\po_l}\tr\, (h_{f_l}\Pi_l) & \approx -2 \sum_{\m=0}^3 \s_{\m l} a_\m \left( 
\cos\f{\g\Th_\m -\chi_\m}2 \sinh \f{\Th_\m}2 + i \sin\f{\g\Th_\m -\chi_\m}2 \cosh \f{\Th_\m}2  \right).
\end{align}\end{subequations}

Let us look first at the Hamiltonian constraint. $\im\tr\,h_f=0$ is solved by $\Th_\m=0$ or $\g\Th_\m=\chi_\m$, which in turn imply
$\Xi_i =0$, or
\be\label{gXixi}
 \g\Xi_i=\xi^i_{jk},
\ee
where the labels $jk$ in the right-hand side are fixed by the face we are looking at, see \Ref{chixi}. The first solution corresponds to a degenerate configuration with zero extrinsic curvature.\footnote{To be more precise, the solution should read $\Xi_{i}=2ik \pi, \; k\in \Z$. As explained in \cite{Sorkin:1975ah}, the dihedral angle of a Lorentzian 4-simplex are defined with shifts of $i\pi/2$ when the vctors belong to two adjacent Lorentz quadrants. This allows compatibility between degenerate configurations and the closure constraint.} 
The second solution, allowing for non-zero extrinsic curvature and expressing it in terms of 3-dimensional geometry, is the one we are interested in. First of all, notice that the relation between 4-dimensional and 3-dimensional geometry depends on the Immirzi parameter, as in the continuum theory. Secondly, the solution implies the shape-matching conditions: as the same link $i$ is shared by different faces, $\xi^i_{jk}$ should be independent of $(jk)$ which as we recalled above requires the matching of shapes.

Next, let us look at the $\psi_{lf}$, independently at first of the Hamiltonian constraint.
Recalling their definition \Ref{sec} and that of $\s_{\m l}$, the three secondary constraints on a given face have the structure
\begin{align}
& a_0 C_0 + a_1 C_1 - a_2 C_2 - a_3 C_3 = 0, \nonumber \\
& a_0 C_0 - a_1 C_1 + a_2 C_2 - a_3 C_3 = 0, \nonumber \\
& a_0 C_0 - a_1 C_1 - a_2 C_2 + a_3 C_3 = 0, \nonumber 
\end{align}
where
\be
C_\m = \g \cos\f{\g\Th_\m -\chi_\m}2 \sinh \f{\Th_\m}2 + \sin\f{\g\Th_\m -\chi_\m}2 \cosh \f{\Th_\m}2. 
\ee
Taking sums of three different pairs, we can recast them in the following equivalent form, 
\be\label{aC}
a_0 C_0 = a_i C_i \qquad \forall i=1,2,3.
\ee
Next, we take the ratio
\be\label{malmousque}
\f{a_0 a_i}{a_j a_k} = \f{C_j C_k}{C_0 C_i},
\ee
and observe from \Ref{amu} that
\be\label{uno}
\f{a_0 a_i}{a_j a_k} = \cot^2 \f{\phi_{jk}}2 = \f{1+\cos\phi_{jk}}{1-\cos\phi_{jk}}.
\ee
On the other hand, the right hand side of \Ref{malmousque} can not be easily decomposed in terms of a single trigonometric function because of the $\g$ factors. Partially expanding the expression in terms of the fundamental $\Xi_i$ and $\xi^i_{jk}$ angles, and equating to \Ref{uno}, we obtain
\be\label{due}
\f{1+\cos\phi_{jk}}{1-\cos\phi_{jk}} = \f{1+ f_1(i,j,k)}{1- f_2(i,j,k)},
\ee
with
\begin{align}
&f_1(i,j,k) = -\frac{\cosh (\Xi_j - \Xi_k)}{\cosh \Xi_i}\left[1+\frac{\cos \left[(\g \Xi_j - \xi^j_{ik}) - (\g \Xi_k - \xi^k_{ij})\right]}{\cos (\g \Xi_i - \xi^i_{jk})}\right] +  \frac{\cos \left[(\g \Xi_j - \xi^j_{ik}) - (\g \Xi_k - \xi^k_{ij})\right]}{\cos (\g \Xi_i - \xi^i_{jk})}  
\nonumber \\
& \qquad \qquad + \frac{4}{\gamma} \frac{\sin (\g \Th_j - \chi_j) \cos (\g \Th_k - \chi_k) \cosh \Th_j \sinh \Th_k + \cos (\g \Th_j - \chi_j) \sin (\g \Th_k - \chi_k) \sinh \Th_j \cosh \Th_k }{\cosh \Xi_i \cos (\g \Xi_i - \xi^i_{jk})} + \nonumber \\
& \qquad \qquad  + \frac{4}{\gamma^2} \frac{\sin (\g \Th_j - \chi_j) \cosh \Th_j + \sin (\g \Xi_k - \chi_k) \cosh \Th_k}{\cosh \Xi_i \cos (\g \Xi_i - \xi^i_{jk})},
\end{align}
\begin{align}
&f_2(i,j,k) =  \frac{\cosh (\Xi_j + \Xi_k)}{\cosh \Xi_i}
\left[1+\frac{\cos \left[(\g \Xi_j - \xi^j_{ik}) + (\g \Xi_k - \xi^k_{ij})\right]}{\cos (\g \Xi_i - \xi^i_{jk})} \right] 
-  \frac{\cos \left[(\g \Xi_j - \xi^j_{ik}) + (\g \Xi_k - \xi^k_{ij})\right]}{\cos \g \Xi_i - \xi^i_{jk}} 
\nonumber \\
&  \qquad \qquad - \frac{4}{\gamma} \frac{\sin (\g \Th_0 - \chi_0) \cos (\g \Th_i - \chi_i) \cosh \Th_0 \sinh \Th_i + \cos (\g \Th_0 - \chi_0) \sin (\g \Th_i - \chi_i) \sinh \Th_0 \cosh \Th_i }{\cosh \Xi_i \cos (\g \Xi_i - \xi^i_{jk})} + \nonumber \\
&  \qquad \qquad  - \frac{4}{\gamma^2} \frac{\sin (\g \Th_0 - \chi_0) \cosh \Th_0 + \sin (\g \Th_i - \chi_i) \cosh \Th_i}{\cosh \Xi_i \cos (\g \Xi_i - \xi^i_{jk})}.
\end{align}
The key observation at this point is that $f_1(i,j,k)$ differs in general from $f_2(i,j,k)$, thus \Ref{due} and the secondary constraints have no solutions. This over-constraining of the system stems precisely from the fact that there are more secondary than primary simplicity constraints. By direct inspection, we see that the only configurations for which $f_1(i,j,k) = f_2(i,j,k)$ are those satisfying  $\g \Xi_i = \xi^i_{jk}$. Therefore, the existence of solutions to our secondary constraints requires the matching of shapes, just as above it was the case for the existence of non-degenerate solutions to the Hamiltonian constraint. 
For shape-matched configurations, only the first terms of $f_1$ and $f_2$ survive, and we get
\be\label{tre}
\f{C_0 C_i}{C_j C_k} = \f{\sinh \f{\Th_j}2 \sinh \f{\Th_k}2}{\sinh \f{\Th_0}2\sinh \f{\Th_i}2} = 
\f{\cosh\Xi_j \cosh\Xi_k - \sinh \Xi_j \sinh \Xi_k - \cosh\Xi_i}{\cosh\Xi_i - \cosh\Xi_j \cosh\Xi_k - \sinh \Xi_j \sinh \Xi_k}.
\ee
Hence, equating \Ref{tre} and \Ref{uno}, we arrive at
\be\label{cosinelaw}
\cos\phi_{ij} = \f{\cosh\Xi_k - \cosh\Xi_i \cosh\Xi_j}{\sinh\Xi_{i} \sinh\Xi_j}.
\ee
These are nothing but the spherical cosine laws in the Lorentzian case (see e.g. \cite{Sorkin:1975ah}),
expressing the 3-dimensional dihedral angles in terms of 4-dimensional ones.

When the graph is the boundary of a 4-simplex, all faces are 3-valent, and these formulas can be immediately inverted, 
to express the extrinsic curvature in terms of the 3-dimensional data,
\be\label{final}
\cosh\Xi_{i} = \f{\cos\phi_{jk} + \cos\phi_{ij} \cos\phi_{ik}}{\sin\phi_{ij} \sin\phi_{ik}}.
\ee
Because of the shape matching conditions imposed by the first half of the solution, \Ref{gXixi}, we are dealing effectively with a Regge geometry, and plugging \Ref{final} in the expression for the reduced link holonomies $h_l$, we obtain precisely Regge's discrete version of the Levi-Civita compatibility of the connection with the flat metric of the 4-simplex. 

Equations \Ref{gXixi} and \Ref{cosinelaw}, the solutions to the secondary constraints, are the main result of this paper. They show that our dynamically generated secondary constraints imply a Levi-Civita condition on the holonomy, as well as the matching of shapes.
As desired, the procedure provides an explicit gauge fixing of the diagonal simplicity constraint's orbits via \Ref{cosinelaw}, and tells us how the extrinsic geometry can be recovered from purely SU(2) data: on-shell of the constraints, the SU(2) angle $\xi^i_{jk}$ equals $\gamma$ times the 4-dimensional dihedral angle, in turn determined by \Ref{final}. Notice that this perfectly reproduces the result previously found by simply evaluating holonomies and fluxes on a Regge geometry \cite{IoCarloGraph}.

Let us also discuss the case of more general graphs. Consider first the case of a graph dual to an arbitrary triangulation. Unlike for the 4-simplex, which only had tri-valent faces, there will be now faces of arbitrary valence $p$. The Hamiltonian and secondary constraints still have the structure \Ref{redconstraints}, but now $p$ products of cosines and sines appear in the $a_i$ coefficients, and $p$ summations over angles in the $\Th_i$ and $\chi_i$, making the relation between 3-dimensional and 4-dimensional angles more complicated than the spherical cosine laws \Ref{cosinelaw}. It is to be expected that manipulations similar to those performed above lead to the angle relations for an arbitrary flat polytope, holding only if shapes match, but we have not tried to prove it. In fact, the simplest thing to do is to split the higher valence faces into tri-valent ones introducing a virtual link, and reducing the relation between 3-dimensional and 4-dimensional angles to the simplicial one. The same procedure applies also to graphs dual to general cellular decompositions, but with additional difficulties. First of all, one needs to reconstruct first the shape-matched polyhedra, to make sure the various angles are proper dihedral angles. Second, the number of independent angles $\phi_{ij}$, that is $\sum_n 2(f_n - 3) = 4L-6N$ for a closed graph, is larger that the number $L$ of $\Xi_i$. Indeed, the chopping procedure introduces extra dihedral angles $\Xi_i$ between tetrahedra inside the same polyhedron, which correspond to the possible additional extrinsic curvature allowed by the finer triangulation, and which are fixed by working with the coarser cellular decomposition.

%%----------------------------------------------------------------------------
\section{Spherical cosine laws from flatness}
%%----------------------------------------------------------------------------

As we pointed out in Section 3, flatness implies makes the secondary constraints redundant, see \Ref{sec}. Now that we have exposed the geometrical meaning of the secondary constraints as spherical cosine laws and shape matchings, it follows that it should be possible to derive the latter directly from the flatness condition. Here we provide a simple proof of this statement, based on the twisted geometry parametrisation. Our result extend to 4-dimensions the 3-dimensional geometric interpretation of the vertex condition derived by 't Hooft in his polygon model \cite{'tHooft}.

For simplicity, we consider again a 3-valent face, with links $i$, $j$ and $k$. The flatness condition is
\begin{align}
&h_f = h_i h_j h_k = \Id,
\end{align}
which, in the present Lorentzian case, is a complex equation.
Using the spinorial representation (\ref{hf-red}) on shell of the simplicity constraints,
this equation can be rewritten as
\begin{align}
& T_{11} \Ket{\tilde{z}_i} \Bra{z_k} + T_{12}  \Ket{\tilde{z}_i} \BraQ{z_k} + T_{21} \KetQ{\tilde{z}_i} \Bra{z_k} + T_{22} \KetQ{\tilde{z}_i} \BraQ{z_k} \stackrel{!}{=}\Id \left(||z_k||^2 ||z_j||^2 ||z_i||^2\right),
\end{align}\label{eq:Hooft}
where
\begin{subequations}\label{Ts}
\begin{align}
&T_{11} = e^{-\eta_i - \eta_k} \Big( e^{-\eta_j} \Scal{z_i}{\tilde{z}_j} \Scal{z_j}{\tilde{z}_k} + e^{\eta_j} \MissQ{z_i}{\tilde{z}_j} \Miss{z_j}{\tilde{z}_k}\Big)\\
&T_{12} = e^{-\eta_i + \eta_k} \Big( e^{-\eta_j} \Scal{z_i}{\tilde{z}_j} \MissQ{z_j}{\tilde{z}_k} + e^{\eta_j} \MissQ{z_i}{\tilde{z}_j} \ScalQ{z_j}{\tilde{z}_k}\Big)\\
&T_{21} = e^{\eta_i - \eta_k} \Big( e^{-\eta_j} \Miss{z_i}{\tilde{z}_j} \Scal{z_j}{\tilde{z}_k} + e^{\eta_j} \ScalQ{z_i}{\tilde{z}_j} \Miss{z_j}{\tilde{z}_k}\Big)\\
&T_{22} = e^{\eta_i + \eta_k} \Big( e^{-\eta_j} \Miss{z_i}{\tilde{z}_j} \MissQ{z_j}{\tilde{z}_k} + e^{\eta_j} \ScalQ{z_i}{\tilde{z}_j} \ScalQ{z_j}{\tilde{z}_k}\Big)
\end{align}
\end{subequations}
and  $\eta_i = \f12(1+ i \gamma) \Xi_i$.
Contracting the matricial equation (\ref{eq:Hooft}) with different vectors, and exploiting the identity $[z_i\ket{z_i}\equiv 0$, we obtain the equations
\begin{align}
T_{11} = \Scal{\tilde{z}_i}{z_k} ||z_j||^2, \quad T_{12} = \MissQ{\tilde{z}_i}{z_k}||z_j||^2,  
\quad T_{21} = \Miss{\tilde{z}_i}{z_k}||z_j||^2, \quad T_{22} = \ScalQ{\tilde{z}_i}{z_k}||z_j||^2,
\end{align}
from which it follows that
\be
T_{11}T_{22} + T_{12}T_{21} = ||z_j||^4 \,\left( |\Scal{\tilde{z}_i}{z_k}|^2 - |\Miss{\tilde{z}_i}{z_k} |^2 \right) 
= ||z_k||^2 ||z_i||^2 ||z_j||^4 \cos \phi_{\tilde{\imath}k},
\ee
where in the last equality we used \Ref{z-geo}.
On the other hand, from the definitions \Ref{Ts} in terms of the nodal invariants 
$E_{i\tl\jmath}= \Scal{z_i}{\tilde{z}_j}, F_{j\tl k} = \Miss{z_j}{\tilde{z}_k}$, we get
\begin{align}
T_{11} T_{22}  + T_{12} T_{21} &= 2 \left( e^{-2 \eta_j} E_{i\tl\jmath} E_{j\tl k} \bar{F}_{\tl \jmath i} \bar{F}_{j\tl k} 
+ e^{2 \eta_j} \bar E_{i\tl\jmath} \bar E_{j\tl k} {F}_{\tl \jmath i} {F}_{j\tl k} \right) + 
(|E_{i\tl\jmath}|^2 - |F_{\tl \jmath i}|^2) (| E_{j\tl k}|^2 - |F_{j\tl k}|^2) 
 \nonumber \\
&= ||z_k||^2 ||z_i||^2 ||z_j||^4 \left(\sin \phi_{\tilde{\jmath}i} \sin \phi_{j\tilde{k}} \cosh \left( 2 \eta_j - i \xi^j_{i\tilde{k}}\right) 
+ \cos \phi_{\tilde{\jmath}i} \cos \phi_{j\tilde{k}} \right),
\end{align}
again using \Ref{z-geo}.
Equating these two expressions, and recalling the definition of $\eta_i$, we get
\begin{align}
&\sin \phi_{\tilde{\jmath}i} \sin \phi_{j\tilde{k}} \cosh \left( \Xi_j + i (\g \Xi_j - \xi^j_{i\tilde{k}}) \right) 
+ \cos \phi_{\tilde{\jmath}i} \cos \phi_{j\tilde{k}}  = \cos \phi_{\tilde{\imath}k}.
\end{align}
The imaginary part of this equation gives the shape matching conditions, and the real part, the spherical cosine law. 
Twisted geometries elegantly show how both conditions derive from requiring a flat Lorentzian holonomy.

%----------------------------------------------------------------------------
\section{Conclusions}
%----------------------------------------------------------------------------

In this paper we addressed the question of the presence and meaning of secondary simplicity constraints in a model in which they can be dynamically derived on a fixed graph.
We answered both questions posed in the introduction in the affirmative: firstly, dynamical stabilisation of the primary simplicity constraints leads to secondary constraints; secondly, these are a discrete version of the Levi-Civita condition for the connection, in the form of the spherical cosine law between 4d and 3d dihedral angles, and their solution requires the shape-matching conditions restricting a twisted geometry to a Regge geometry. In our construction, the origin of the additional shape-matching conditions is a direct consequence of the different locality properties of the constraint used, on the links of the graphs the primary simplicity, on the faces the flatness Hamiltonian constraint. 
As a consequence, each link has as many secondary constraints as there are faces sharing it.
In explicitly solving the constraints, we used heavily the parametrization in terms of twisted geometries, and especially the results of \cite{FreidelJeff13} that introduced a geometric characterisation of the SU(2) invariants in terms of framed polyhedra \cite{EteraSU2UN}, and a gauge-independent version of the twist angle $\xi$. In particular, our results show in what precise sense this angle carries the extrinsic geometry, and the presence of the Immirzi parameter in the relation.

An immediate question for future research is whether our analysis could be carried through also with a Hamiltonian constraint allowing for curvature. One could for instance think of the formalism of quasi-constraints of \cite{DittrichHoehn, Dittrich:2013jaa, Hoehn:2014aoa}, or of attempts at discretising the Hamiltonian constraint of LQG on a fixed graph such as \cite{Immirzi94, Zapata:1996ij}. 
Let us also point out the model by Wieland \cite{WielandHamSF, WielandNew}, where a continuous time parameter is introduced while working with a discrete spacetime triangulation proper to spin foams. It would be interesting to add the Hamiltonian constraint here considered, or a more general one, to that model, to study the possible appearance of secondary constraints.

%----------------------------------------------------------------------------
\subsection*{Acknowledgements}
%----------------------------------------------------------------------------
We thank Sergei Alexandrov, Bianca Dittrich, Pietro Don\'a and Wolfgang Wieland for discussions and comments on a draft of the paper.

\appendix
%%----------------------------------------------------------------------------
\section{Canonical Analysis}
%%----------------------------------------------------------------------------

The primary Hamiltonian of our system is
\be
H = \sum_f N_f \hh_f  + \sum_l \left(\n_l C_l +\bar \n_l \bar C_l + \l_l S_l +\m_l F_l +\bar \m_l \bar F_l + \tl\m_l \tl F_l +\bar{\tl \m}_l \bar{\tl F}_l \right) +\sum_{n} \vec n_n \cdot \vec{G}_n\, .
\ee
Since all constraints are invariant under the area matching and SU(2) transformations, we can drop these two constraints from the analysis. The only non-vanishing brackets on shell of the constraints are $\{F_l,\bar F_l\}$ and $\{\hh_f, S_l\}$, given respectively by \Ref{FFbar} and \Ref{HD}. The first give equations fixing the Lagrange multipliers $\m_l$ and $\tl\m_l$. The second enter the stability of the primary simplicity constraints, which gives
\be\label{AStabS}
\dot S_l 
\approx \sum_{f} N_f \{\hh_f, S_l\}
= 2 \sum_{f| l\in\p f} N_f  (\g\re +\im)\left[\f1{\po_l}\tr(h_{f_l}\Pi_l)\right]  = 0.
\ee
Requiring as customary as non-vanishing lapse, \Ref{AStabS} gives the secondary constraints
\be
\psi_{lf} := \{\hh_f, S_l\} \approx 2 (\g\re +\im)\left[\f1{\po_l}\tr(h_f\Pi_l)\right] \approx N_l^i (\g\re +\im)\tr(h_f\s_i),
\ee
where $\approx$ means on shell of the primary simplicity constraints, so $h_f$ here equals the product of reduced link elements \Ref{hf-red}.
The secondary constraints provide, as in the continuum, equations relating holonomies and intrinsic geometry. Their explicit solution is studied in the main text, and it will not be needed here. Notice that imposing both $\hh_f=\psi_{lf}=0$ does not imply full flatness; a direct calculation shows that $\re\tr h_f$ is still arbitrary.
Following Dirac's procedure, we add the secondary constraints to the action, with Lagrange multipliers $M_{lf}$. Imposing the stability of the complete set of constraints, we arrive at the following system,
\begin{align}\label{stabS}
&\dot{S}_{l} \approx  \sum_{f',l'} M_{l'f'} \Pois{\psi_{l'f'}}{S_{l}}, \\ \label{stabF}
&\dot{F}_{l} \approx  2i\,j_{l}\bar{\mu}_{l}  +  \sum_{f} N_f \{\hh_f, F_{l}\} + \sum_{f',l'} M_{l'f'} \Pois{\psi_{l'f'}}{F_{l}}, \\ 
&\dot{\tl F}_{l} \approx  2i\,j_{l}\bar{\tl\mu}_{l}  +  \sum_{f} N_f \{\hh_f, \tl F_{l}\} + \sum_{f',l'} M_{l'f'} \Pois{\psi_{l'f'}}{\tl F_{l}}, 
\label{stabFtl} \\ 
\label{stabH}
&\dot{\hh}_f \approx \sum_l \left[\m_l\{F_l, \hh_f\} + \tl\m_l\{\tl F_l, \hh_f\} + {cc} \right]+ \sum_{l'f'} M_{l'f'}\{\psi_{l'f'}, \hh_f\}, \\ \label{stabpsi}
&\dot{\psi}_{lf} \approx \sum_{f'} N_{f'} \{\hh_{f'}, \psi_{lf} \} + 
\sum_{l'} \lambda_{l'} \Pois{S_{l'}}{\psi_{lf}} + 
\left[ \m_l \Pois{F_{l'}}{\psi_{lf}}  + \tl\m_l \Pois{\tilde{F}_{l'}}{\psi_{lf}} +cc \right]
+ \sum_{lf} M_{lf} \{\psi_{lf}, \psi_{l'f'} \}.
\end{align}
The sums here should be understood always as restricted to intersecting links and faces, as otherwise the spinorial variable commute. The non-vanishing brackets appearing above can be explicitly evaluated, giving
\begin{align}\label{Spsi}
& \{ S_l, \psi_{l'f} \} \approx \d_{l'l} \, \f\g2 \re\tr h_f 
+ \d_{l'\neq l} \, \left(2\g \re +(1-\g^2)\im\right) \left[\f4{\po_l\po_{l'}} \tr\, \left(h_{f_l} \Pi_{l'}^{(l)}\Pi_l \right)\right] \\
& \{\hh_f, \psi_{lf'}\} \approx \f1{4j_l} \Big[ \sin\th \big(\re\tr h_f \re\tr h_{f'} - 2 \re\tr h_{f\circ {f'}^{-1}} \big) 
+ 2\cos\th \, \im\tr h_{f\circ {f'}^{-1}} \Big] \\
& \{ \hh_f, F_l \} \approx \f i2 \Big[ \f1{\po_l} \bra{\om_l} h_{f_l} \ket{\pi_l} + \f1{\overline{\po}_l} \bra{\om_l} h_{f_l}^\dagger \ket{\pi_l} \Big] 
\\ \label{psiF}
& \{ \psi_{l'f}, F_l \} \approx \d_{l'l} \left[ \f{\g-i}4 \f1{\po_l} \bra{\om_l} h_f \ket{\pi_l}
+ \f{\g+i}4 \f1{\overline{\po}_l} \bra{\om_l} h_f{}^\dagger \ket{\pi_l} \right] \\
& \qquad \qquad \; + \d_{l'\neq l} \left[-\f{\g-i}{\po_l\po_{l'}} \bra{\om_l} h_{f_l}\Pi_{l'}^{({l})} \ket{\pi_l}
+\f{\g+i}{\overline{\po}_l\overline{\po}_{l'}} \bra{\om_l} \big(h_{f_l}\Pi_{l'}^{({l})}\big)^\dagger \ket{\pi_l}\right] 
\nonumber \\ \label{psipsi}
& \Pois{\psi_{\phi t}}{\psi_{fl}} =  \left[2(\gamma^2-1) \mathrm{Re} + 4 \gamma \mathrm{Im} \right] \left(  \Pois{\frac{\tr(h_{\phi} \Pi_t)}{\po_t}}{\frac{\tr(h_f \Pi_l)}{\po_l}} \right),
\end{align}
where
\begin{align}
\Pois{\frac{\tr(h_{\phi} \Pi_t)}{\po_t}}{\frac{\tr(h_f \Pi_l)}{\po_l}} & =
\delta_{lt} \left[ \mathrm{Tr} (h_{\phi_t} \vec{\tau}) \times \mathrm{Tr} (h_{f_l} \vec{\tau}) \right] \cdot \vec{\Pi}_l 
\\ \nonumber 
&+ \delta_{t\neq l,\partial f} \left[ \f{4}{\po_t} \, \mathrm{Tr} (h_{f_l}\Pi_t^{(l)} \Pi_l) -\f12 \tr (h_{f_{t+1}}\Pi_l^{(t+1)}h_{\phi_t})
-\f14 \tr h_\phi \tr h_{f_l}\Pi_l \right]
\\ \nonumber 
&- \delta_{l\neq t,\partial \phi} \left[ \f{4}{\po_l} \, \mathrm{Tr} (h_{\phi_t}\Pi_l^{(t)} \Pi_t) -\f12 \tr (h_{\phi_{l+1}}\Pi_t^{(l+1)}h_{f_l})
-\f14 \tr h_f \tr h_{\phi_t}\Pi_t \right]
\end{align}
and $\circ$ means path composition.

The geometric interpretation of some of the coefficients can be made more explicit in terms of curvature and face normals by expressing the Lorentzian spinors in terms of the SU(2) spinors and associated fluxes. 
To that end, it is convenient to parametrise the face holonomy as
\be
h_{f_l} = \cos \a_{f_l} \Id + i \sin\a_{f_l} \, \hat u_{f_l}\cdot \vec\s,
\ee
where $\hat u$ is a complex unit vector and $\a\in\R$ on-shell of the Hamiltonian.
Then, the brackets read
\begin{align}\label{Spsi1}
& \{ S_l, \psi_{l'f} \} \approx \g\cos\a_{f_l} \left(\d_{l'l} + \d_{l'\neq l} \, 4 \vec N_{l} \cdot \vec N_{l'}^{(l)} \right)
-2  \d_{l'\neq l} \,  \sin\a_{f_l} \, \left(2\g \re +(1-\g^2)\im\right) \left(\hat u_{f_l} \right) \cdot\vec N_{l} \times \vec N_{l'}^{(l)} \\
& \{ \hh_f, F_l \} \approx \f{e^{i\f\th2}}{2||\om||^{2i\g}} \sin\a_{f_l} \im\left(e^{-i\f\th2} \hat u_{f_l} \right)
\cdot (\vec F_l+i \vec F_l\times\vec N_l)
\\ \label{psiF1}
& \{ \psi_{l'f}, F_l \} \approx  \f{1-i\g}{||\om||^{2i\g}} 
 \bigg( \d_{l'l} \, \f14  \sin\a_{f_l} \im\left(e^{-i\th} \hat u_{f_l} \right) 
\\\nonumber & \hspace{3cm}  - \d_{l'\neq l} \, 
\left[\sin\th\cos\a_{f_l} \vec X_{l'}^{(l)} -\sin\a_{f_l} \vec X_{l'}^{(l)} \times \im\left(e^{-i\th} \hat u_{f_l} \right)\right]\bigg)
\cdot (\vec F_l+i \vec F_l\times\vec N_l) \\ \label{psipsi1}
& \Pois{\psi_{\phi t}}{\psi_{fl}} \approx \left[2(\gamma^2-1) \mathrm{Re} + 4 \gamma \mathrm{Im} \right] \Bigg\{
\delta_{lt} \, (i\g-1)\sin \alpha_{\phi_l} \sin \alpha_{f_l} \, \left(\hat{u}_{\phi_l} \times \hat{u}_{f_l} \right)\cdot \vec{X}_l 
\\ \nonumber 
&\hspace{2cm} +\delta_{t\neq l,\partial f}\,\f{i\g-1}2\Bigg[ 4i \left[ \sin \alpha_{f_l} \left(\vec{N}_t^{(l)} \times \vec{X}_l\right) \cdot \hat{u}_{f_l} -\cos \alpha_{f_l} \vec{N}_t^{(l)} \cdot \vec{X}_l \right] - \cos \alpha_{\phi} \sin \alpha_{f_l} \, \hat{u}_{f_l} \cdot \vec{X}_l \\
& - \cos \alpha_{\phi_t} \sin \alpha_{f_{t+1}} \hat{u}_{f_{t+1}} \cdot \vec{X}_l^{(t+1)} - \sin \alpha_{\phi_t} \cos \alpha_{f_{t+1}} \hat{u}_{\phi_{t}} \cdot \vec{X}_l^{(t+1)} + \sin \alpha_{\phi_t} \sin \alpha_{f_{t+1}}  \left(\hat{u}_{\phi_t} \times \hat{u}_{f_{t+1}} \right) \cdot \vec{X}_l^{(t+1)} \Bigg] \nonumber 
\\ \nonumber 
& \hspace{2cm} -\d_{l\neq t,\partial \phi} \,\f{i\g-1}2 \Bigg[ 4i \left[ \sin \alpha_{\phi_t} \left(\vec{N}_l^{(t)} \times \vec{X}_t\right) \cdot \hat{u}_{\phi_t} -\cos \alpha_{\phi_t} \vec{N}_l^{(t)} \cdot \vec{X}_t \right] - \cos \alpha_{f} \sin \alpha_{\phi_t} \, \hat{u}_{\phi_t} \cdot \vec{X}_t \\
& - \cos \alpha_{f_l} \sin \alpha_{\phi_{l+1}} \hat{u}_{\phi_{l+1}} \cdot \vec{X}_t^{(l+1)} - \sin \alpha_{f_l} \cos \alpha_{\phi_{l+1}} \hat{u}_{f_{l}} \cdot \vec{X}_t^{(l+1)} + \sin \alpha_{f_l} \sin \alpha_{\phi_{l+1}}  \left(\hat{u}_{f_l} \times \hat{u}_{\phi_{l+1}} \right) \cdot \vec{X}_t^{(l+1)} \Bigg]\Bigg\}  \nonumber
\end{align}

Solving the constraint stabilisation system goes as follows. First of all, \Ref{stabF} and \Ref{stabFtl} fix the Lagrange multipliers $\m_l$ and $\tl \m_l$ as linear combinations of $N_f$ and $M_{lf}$. Then, from \Ref{Spsi1} and \Ref{psipsi1}, we see that $L$ secondary constraints are second class with the primary $S_l$, and the remaining are second class among themselves. Accordingly, \Ref{stabS} fixes $L$ of the multipliers $M_{lf}$, and \Ref{stabpsi} fixes the $L$ multipliers $\l_l$ plus the remaining $M_{lf}$. All these multipliers are thus given by linear combinations of the lapses $N_f$, with coefficients which are functions on the phase space.
Using these results, the stability of the Hamiltonian constraint \Ref{stabH} reads
\be
\sum_{f'} N_{f'} C_{ff'} \stackrel{!}{=} 0,
\ee
which gives rise to a tertiary constraint 
\be\label{tertiary}
 C_{ff'} =0. 
 \ee
Because of the discrete nature of the system, this constraint is rather complicated, in particular as it involves the inverse of the matrix \Ref{psipsi1}. Fortunately, its explicit expression is not needed here, as a simple counting shows that the system is already over-constrained. To see that, notice that at this point all constraints are second class, except for $\vec G_n$, which are first class, and $\hh_f$ and $C_{ff'}$, which are yet to be classified. Ignoring for the time being the possible orbits of $\hh_f$ and $C_{ff'}$ to be removed, the counting gives
\be
{\dim \G} = 12L - 6N -5L -F- \#_\psi - \#_C,
\ee
where the number of secondary constraints $\#_\psi$ is given by all possible independent pairs $lf$, and 
the number of tertiary constraints $\#_C$ by all possible pairs $ff'$ sharing a link.
In the case of a triangulation, $L=2N$, and furthermore 
$\#_\psi = \#_C = 3L$, giving dim\,$\G=-2L-F<0$. Similarly for an arbitrary graph dual to a cellular decomposition, where $\#_\psi$ and $\#_C$ grow typically faster with $L$.

Therefore, the system including the tertiary constraints is overdetermined and will admit in general no solutions. There is however a special case, when the whole system is stable, which is given by the flat solution
\be
h_f = \Id \quad \forall f.
\ee 
In this case, the only non-zero brackets are
\begin{align}
& \{ S_l, \psi_{l'f} \} \approx \d_{l'l} \, \g + 4\g  \, \d_{l'\neq l} \vec N_l \cdot \vec N_{l'}^{(l)}, \\
& \{ \psi_{l'f}, F_l \} \approx \d_{l'\neq l} \, \f{i\g-1}{||\om||^{2i\g}} \sin\th \, \vec X_{l'}^{(l)} \cdot (\vec F_l+i \vec F_l\times\vec N_l), \\
& \Pois{\psi_{\phi t}}{\psi_{fl}} \approx  4\g \left[ \delta_{t,\partial f} \, j_l \vec N_l \cdot \vec N_{t}^{(l)}
-  \delta_{l,\partial \phi} \, j_t \vec N_t \cdot \vec N_{l}^{(t)} \right],
\end{align}
and the system collapses to
\begin{align}\label{stabS0}
&\dot{S}_{l} \approx  -\g\sum_{f|l\in\p f}M_{lf} + \sum_{f',l'\neq l} M_{l'f'} \Pois{\psi_{l'f'}}{S_{l}}, \\ \label{stabF0}
&\dot{F}_{l} \approx  2i\,j_{l}\bar{\mu}_{l}   + \sum_{f',l'\neq l} M_{l'f'} \Pois{\psi_{l'f'}}{F_{l}}, \\ \label{stabFtl0}
&\dot{\tl F}_{l} \approx  2i\,j_{l}\bar{\tl\mu}_{l}  + \sum_{f',l'\neq l} M_{l'f'} \Pois{\psi_{l'f'}}{\tl F_{l}}, \\ 
\label{stabH0}
&\dot{\hh}_f \approx 0, \\ \label{stabPsi0}
&\dot{\psi}_{lf} \approx  
\sum_{l'} \lambda_{l'} \Pois{S_{l'}}{\psi_{lf}} + 
\left[ \m_l \Pois{F_{l'}}{\psi_{lf}}  + \tl\m_l \Pois{\tilde{F}_{l'}}{\psi_{lf}} +cc \right]
+ \sum_{lf} M_{lf} \{\psi_{lf}, \psi_{l'f'} \}
\end{align}
This system is immediately stable, with the same structure exposed above; \Ref{stabS0} and \Ref{stabPsi0} determine the $\l_l$ and $M_{lf}$ multipliers, \Ref{stabF0} and \Ref{stabFtl0} the $\mu_l$ and $\tl\mu_l$, and the Hamiltonian is first class.
Therefore, although our starting Hamiltonian was not imposing full flatness, it arises, as well as the secondary constraints and associated shape matching conditions, as solution from the stability procedure. Finally, notice that the orbits of the Hamiltonian contraint are in one-to-one correspondence with the edge lengths of the cellular decomposition, suggesting that we have freedom in varying them while keeping a flat bulk. While this matches with the result of canonical Regge calculus for a 4-simplex (e.g. \cite{DittrichRyan}), it is an interesting question to compare the situation in more general cases, that we leave open to future research.

%----------------------------------------------------------------------------
\providecommand{\href}[2]{#2}\begingroup\raggedright\endgroup

%----------------------------------------------------------------------------
\end{document}